\documentclass[graybox]{svmult}

\usepackage{mathptmx}       
\usepackage{helvet}         
\usepackage{courier}        
\usepackage{type1cm}        
\usepackage{makeidx}         
\usepackage{graphicx}        
\usepackage{multicol}        
\usepackage[bottom]{footmisc}

\makeindex             

\begin{document}

\title*{Towards degeneracy problem breaking by large scale structures methods}
\author{\'Alvaro de la Cruz Dombriz}
\institute{Astronomy, Cosmology and Gravity Center (ACGC) and
Department of Mathematics - Applied Mathematics,
University of Cape Town, Rondebosch 7700,  South Africa. \\ \email{alvaro.delacruzdombriz@uct.ac.za} 
}
%
%
\maketitle

\abstract{An arguable aspect of the modified gravity theories is that many of them present the so-called {\it degeneracy problem}. For instance, the cosmological evolution, gravitational collapse and the main features of standard black-hole configurations, can be mimicked by many of those theories. In this communication we revise briefly the appropriate observable quantities to be measured in order to discard alternative theories to $\Lambda$CDM, such as the observed growth of scalar perturbations with Sloan data and the CMB tensor perturbations evolution.
}

\section{Introduction}
\label{sec:1}

Modified gravity \cite{review} 
has been shown to be able to mimic both the dark energy (DE) and the inflationary eras \cite{unification}.  However the use of large scale observations, such as Ia type supernova, baryon acoustic oscillations, or the cosmic microwave background (CMB), which only depend upon the expansion history of the Universe is not enough to determine uniquely the nature and the origin of DE. Let us rephrase the argument: 
identical cosmological background evolutions 
can be explained by a pleiad of theories. This is the so-called {\it degeneracy problem}, whose breaking requires measurements 
not only sensitive to the 
cosmological expansion but, for instance, the evolution of scalar perturbations \cite{Perturbations}, the stability of cosmological solutions when subjected to small perturbations \cite{Dombriz-Saez} and the existence of General Relativity (GR)-predicted astrophysical objects such as black holes \cite{BH}. Finally, the study of CMB tensor perturbations may also shed some light about the viability of modified gravity theories \cite{CMB-ERE2011, PRD-CMB-Dombriz, Strings, Branes, Challinor, Zhong:2010ae}.  

In this realm, the simplest and in fact the most studied modification of  the Hilbert-Einstein action is generalized to a general function of the Ricci scalar $R$, dubbed $f(R)$ gravity theories  \cite{Ref5}-\cite{Ref6} whose action can be written as 
>%
 \begin{eqnarray}
 {\cal{A}}=\frac{1}{16\pi G}\int{{\rm d}^{4}x\sqrt{-g}\left(R+f(R)+2{\cal{L}}_{m}  \right) }\;,
 \end{eqnarray}
 where the symbols hold their usual meanings.
In addition to reproducing the entire cosmological history \cite{gr-qc/0607118} and despite some shortcomings  \cite{Ref6}, these theories may behave quite well on local scales, where the GR limit must be recovered \cite{Viable}. As for any alternative theory of gravity, 
in $f(R)$ theories, the density contrast evolution, the CMB perturbations and the backreaction mechanism \cite{Dombriz_Backreaction}, if the latter is assumed to be true, need to be studied 
in order to unveil the potential distinct features of these scenarios. 
In the present investigation we sketch the main 
features and steps to study the two first issues in $f(R)$ theories. 

\section{Scalar perturbations in $f(R)$ theories}
\label{sec:2}
The density contrast evolution for $f(R)$ theories obeys a fourth-order differential equation \cite{PRD_Dombriz_2008}.
The resulting equation for the density contrast $\delta$  can be written as follows:
\begin{eqnarray}
&&\beta_{4,f}\delta^{iv}+\beta_{3,f}\delta'''
+(\alpha_{2,EH}+\beta_{2,f})\delta''
+(\alpha_{1,EH}+\beta_{1,f})\delta'+ 
(\alpha_{0,EH}+\beta_{0,f})\delta \,=\,0\nonumber\\
&&
\label{density_contrast_eqn}
\end{eqnarray}
where the coefficients $\beta_{i,f}$  $(i\,=\,1,...,4)$ 
involve terms that disappear for $f(R)$ functions linear in $R$ (i.e., GR) whereas $\alpha_{i,EH}$  $(i\,=\,0,1,2)$ involve the linear part in $R$ of $f(R)$.
Thus, the quasi-static limit ($k>>\mathcal{H}$) of (\ref{density_contrast_eqn}) becomes  \cite{PRD_Dombriz_2008}
\begin{eqnarray}
&&\delta''+\mathcal{H}\delta'+\frac{(1+f_{R})^{5} \mathcal{H}^{2} 
(-1+\kappa_1)(2\kappa_1-\kappa_2)-\frac{16}{a^8}
 f_{RR}^{4}(\kappa_2-2)k^{8}8\pi G \rho_{0}a^2}
{(1+f_{R})^{5}(-1+\kappa_1)+\frac{24}{a^8}f_{RR}^{4}(1+f_{R})
(\kappa_2-2)k^{8}}\delta \,=\,0\nonumber\\
&&
\label{density_contrast_eqn_QSA}
\end{eqnarray}
Contrarily to its counterpart for $\Lambda$CDM, the 
coefficients in (\ref{density_contrast_eqn}) depend both upon the model under consideration and the wavenumber $k$. This fact gives rise to
$k$-dependent transfer functions that may alter dramatically the matter power spectra \cite{PRD_Dombriz_2008, PRL_Dombriz_2009, JCAP_Dombriz_2013}.
Available  data \cite{SDSS} using luminous red galaxies in the Sloan Digital Sky Survey (SDSS) were able to measure the large-scale real-space power spectrum. 
These measurements were used to sharpen the constraints on cosmological parameters \cite{Tegmark} and may be straightforwardly compared with the predictions made by gravity theories
\cite{PRL_Dombriz_2009, Tsujikawa_BOSS}. Very recently a full study for the $R^n$ models \cite{JCAP_Dombriz_2013} have stressed the importance of the initial conditions in the perturbed equations
which determine the evolution of the transfer function. Consequently, this method provides an excellent arena to impose 
tight constraints for modified gravity models that are claimed to be valid 
once compared with existing and future data \cite{other_collab}.

\section{CMB perturbations in $f(R)$ theories}
\label{sec:2}
The study of the CMB tensor perturbations in modified gravity theories has not received much interest in comparison 
with the scalar counterpart. This fact has laid in the difficulty of obtaining 
the required tensor perturbed equations which are in general of higher order. An alternative route in order to circumvent this difficulty consists of tackling the problem 
by using the simulations performed by several codes available such as CAMB \cite{CAMB_Lewis} 
based upon modifications of CMBFast \cite{CMB-Fast}.

Different attempts were made for several modified gravity scenarios \cite{Strings} 
but  most of the attention was devoted to the study of the tensor perturbations evolution in the brane-world theories context 
\cite{Branes, Challinor}. 
Finally, with regard to $f(R)$ fourth order gravity theories, the only attempts to encapsulate 
 the main features of tensor perturbation were made in \cite{Zhong:2010ae} and more recently in \cite{CMB-ERE2011, PRD-CMB-Dombriz}. 
The authors of the first investigation analyzed the tensor perturbations of 
flat thick domain wall branes in  $f(R)$ gravity. They showed that under the transverse and traceless gauge, the metric perturbations decouple from the perturbation of the background scalar field which generates the brane. Authors in \cite{CMB-ERE2011, PRD-CMB-Dombriz} addressed for the first time in literature the tensor perturbations full calculations for the $f(R)$ gravity theories in the metric formalism and Jordan frame. These general results were applied to $R^n$ models for different values of $n$ 
%
describing the features that may distinguish those models from Concordance model predictions. This implementation proved the importance of considering the correct background when 
alternative theories of gravity are subjected to this kind of analyses since a relevant contribution to the  $c_{l}^{\rm TT}$ and  $c_{l}^{\rm EE}$ CMB coefficients 
comes from the background implementation.

Thus, exclusions tests for $f(R)$ models can be performed since data for $c_{l}^{\rm TT}$ are already available from WMAP \cite{WMAP} once the scalar contribution are also included. With respect to $c_{l}^{\rm EE}$  once Planck \cite{Planck} measurements are ready, some data may be compared with theoretical predictions.


%
\begin{acknowledgement}
The author acknowledges financial support from URC and NRF (South Africa), MICINN (Spain) project numbers FIS2011-23000, FPA2011-27853-C02-01 and Consolider - Ingenio MULTIDARK CSD2009-00064. 
\end{acknowledgement}
%

%
%
%

%
%

\end{document}